# Capturing the symptoms of malicious code in electronic documents by file's entropy signal combined with Machine learning

**Luping Liu[a], Xiaohai He[a,1], Liang Liu[b] , Lingbo Qing[a] , Yong Fang, Jiayong Liu[b]**

[a] College of Electronics and Information Engineering, Sichuan University, Chengdu, 610065， China

[b] College of Cybersecurity, Sichuan University, Chengdu, 610065 China

**Abstract**-Email cyber-attacks based on malicious documents have become the popular techniques in today's sophisticated attacks. In the past, persistent efforts have been made to detect such attacks. But there are still some common defects in the existing methods including unable to capture unknown attacks, high overhead of resource and time, and just can be used to detect specific formats of documents. In this study, a new Framework named ESRMD (Entropy signal Reflects the Malicious document) is proposed, which can detect malicious document based on the entropy distribution of the file. In essence, ESRMD is a machine learning classifier. What makes it distinctive is that it extracts global and structural entropy features from the entropy of the malicious documents rather than the structural data or metadata of the file, enduing it the ability to deal with various document formats and against the parser-confusion and obfuscated attacks. In order to assess the validity of the model, we conducted extensive experiments on a collected dataset with 10381 samples in it, which contains malware (51.47%) and benign (48.53%) samples. The results show that our model can achieve a good performance on the true positive rate, precision and ROC with the value of 96.00%, 96.69% and 99.2% respectively. We also compared ESRMD with some leading antivirus engines and prevalent tools. The results showed that our framework can achieve a better performance compared with these engines and tools.

## 1. Introduction

[1]Corresponding Author: XIAOHAI HE, Email: hxh@scu.edu.cn

Cyber-attack activities aimed at stealing user's credentials or doing some damages have been increasing over the last years. Email remains a vital channel for malware distribution in today's sophisticated and complex threat environment. Attackers often send emails to their targets and encourage them to open the malicious attachments by using social engineering. Mostly, they tend to use non-Executable files such as Microsoft Office and PDF as vectors to deliver malwares.

Email cyber-attacks with malicious documents were very prevalent in the last year according to the Cisco 2018 Annual Cybersecurity report [28].The researchers analyzed the email cyber-attacks from January through September in 2017 and identified some common file extensions in malicious email attachments. Their investigation showed that most of malicious documents (38 percent) are Microsoft Office formats such as Word, PowerPoint, and Excel. While Malicious PDF file extensions rounded out the top three, accounting for nearly 14 percent. Most of the malicious documents rely on exploiting vulnerabilities in their interpreters to launch attacks. Those attacks are more dangerous and usually harder to be detected compared with other attacks such as Macro-based attacks. No interactions with users are needed and malwares can be released silently. Therefore, email cyber-attacks based on software vulnerabilities are prevalent in Advanced Persisted Thread (APT) attacks. From 2017 to 2018, there were numerous critical vulnerabilities and exposures in Microsoft Office published, such as CVE-2017-0199[3], CVE-2017-8759 [5] and CVE-2017-11882 [6]. All of them can be exploited to different versions of Microsoft Office Products and operating systems from Win XP to Win 10. Many APT attacks based on those vulnerabilities had been found in the wild according to the reports from security vendors [4; 9]. Continued attacks utilizing malicious documents make malicious document detection a pressing problem. In recent years, many efforts have been made to address this problem. Generally speaking, these methods can be divided into 2 categories: static-based analysis and dynamic-based analysis.

Static analysis parses the documents and searches the contents of the file based on signature-bytes or pattern features from shellcode to identify whether the documents are clean or not[24]. The signature-based or pattern-matching method become ineffective if the attackers use some techniques (e.g., obfuscation, polymorphism) to generate new variety of malwares and even zero day exploits to create a new malware. Then researchers started to seek machine learning methods to work around this problem. Machine learning methods automatically learn a discriminant mechanism to detect malicious documents. So far, all the existing machine learning methods just focus on extracting features from the structural data or metadata of the file [26; 29; 45; 51; 52]. The machine learning methods based on structural data or metadata are more effective. However, the robustness of these methods against motivated adversaries is not good enough. It has been demonstrated that the methods based on structural features extraction can be evaded in principled and motivated ways [31; 40; 48; 57]. What's more, they can be only used to detect malicious files of certain types: XML-based Office and PDF. Dynamic analysis runs the document in controlled environments and traces the dynamic behavior, such as API calls, control flow of the execution and so on. Whether the dynamic behavior of the target process are normal or not is then estimated based on statistical analysis or designed rule [25; 27; 32; 43; 47; 56]. Dynamic approaches have inevitable drawbacks. Firstly, it costs a lot of time to capture the behavior of the target process, which makes it difficult to be applied to large-scale detection. Secondly, it has a high-overhead of computing resource. In recent years, some researches have started using signal processing approaches to reveal symptoms of malicious code. They aimed at detecting encrypted or compressed malwares based on analyzing the entropy information of the files [19-21; 46; 55], which found that the techniques the malware authors employed, such as obfuscated shellcode, metamorphisms code and frivolous instruction, make the distribution of entropy characteristic. Therefore, they treat the entropy sequence of the file as a non-stationary time series, and try many different methods to extract features from the entropy distribution. Their results show that the patterns hidden in the entropy signal of the malares is helpful to identify them [23; 54; 55].

Inspired by these findings, in this work, we present a novel approach to detect malicious documents based on entropy distribution of the documents combined with machine learning. In our preliminary investigation, we found that all the malicious attachments used to attack are embedded with external malicious code, such as shellcode, ROP chain, NOPs instructions, malware, Decopy (benign document) and so on. These external data are organized in a certain form and embedded in different locations of the documents. As the samples are highly-structured and many benign documents may contain their own pattern. Therefore, we think the external embed malicious code will change the pattern of the entropy distribution in malicious documents and make them different to the benign files significantly. Based on this hypothesis, we calculate the file entropy of the document and also see it as a non-stationary time series. Then we use different approaches to extract global and structural entropy features from the entropy sequence. Specifically, we extract global features from the file entropy by a simple statistical method. Those features reflect the global information of the file. We extracted structural entropy features by using discrete wavelet decomposition and Bag-of-words. Those structural entropy features reflect the detailed and local characteristics of the file entropy. Then a machine learning algorithm is implemented in these features to train a model. The main difference with our work is that most existing machine learning methods mainly based on the structure of the documents, which can be easily bypassed or attacked and have their own limitation. In contrast, our method extracts features from the entropy distribution of the file which make it more robust and adaptive. What's more, unlike the methods employed to packer and malwares detection, our method uses DWT and BOW to extract detail and local features rather than just extracting overall information, which make it more suitable for detecting malicious documents.

We implement a prototype called ESRMD (Entropy signals Reflect the Malicious document), which aims at enhancing the detection of malicious documents. It is a light and fast framework which it just uses a static analysis technique. Therefore, it can be utilized for large scale documents detection. In order to assess our framework, we creat a dataset and perform extensive experiments on it. The results show that the extracted features can help to enhance the ability to differentiate between malicious and benign documents. We carry out a cross-validation experiment on the created dataset and the evaluations demonstrate that the precision and recall can reach to 96.32% and 95.93%, respectively. We also compare our method with some leading anti-virus and prevalent tools. The result shows that our method can also achieve a better performance. In summary, we make the following contributions:

1. We propose a novel approaches which can be utilized to detect malicious documents based on the entropy distribution of the files. The main difference between previous machine learning methods using to detect malicious document is that our method extracts distinctive features from the entropy signal of the file rather than the structural data or meta-data which makes it is more adaptable and robust.
2. We design and implement ESRDM, a framework that leverages the previous techniques to extract features from the documents, and then combine them with machine learning classifier to identify the new documents are clean or not. This framework can be utilized to detect large scale and various formats of malicious documents.
3. We evaluate ESRDM on a collected dataset with 10381 samples. The results demonstrate that ESRDM can achieve a good performance on both accuracy and efficiency. During the comparison with the leading anti-virus engines and the start-of-the-art, ESRDM acquires a better ability.

The reminder of this paper is structured as follows. In Section. 2, previous work in this area is described. In Section. 3, we illustrate our main framework used to detect malicious documents. In Section. 4, we describe the detail of the methods employed to extract features from the time series. In Section. 5, the evaluation and results are provided. In Section. 6, threats to validity are given. In Section.7, we describe the conclusions and the future work.

# 2. Related Work

In this section, we describe previous work which are used to detect malicious documents. The prior related work can be divided into two categories: static-based analysis and dynamic-based analysis.

## 2.1. Static-based analysis

The static-based analysis relies on analyzing the binary content or structural data of the document to determine whether a document is benign or not. These approaches include three categories. The first category is signature-based methods, which determine whether the sample is clean or not with some signatures designed by experts. There are two famous tools which have been used to detect malicious documents based on patterns matching. YARA [18] is a famous framework which can help users to create signatures and patterns for specific malwares. It has been used by various vendors, such as Cuckoo [2] and Virus Total [16]. Another tool is OfficeCat [10], which scans the document and compares it with predefined signatures. This tool sets up a database which contains the signatures from various vulnerabilities. The main disadvantage of those methods is that they cannot capture zero-day attacks or variant attacks. What's more, they also need to maintain a feature database, so the overhead will become more and more expensive with the increase of the samples.

The second category is based on file content analysis. Those methods parse the document according to the file formats and extract data stream or objects for further analysis. Li *et al*. [37] detects malicious documents by using static content analysis. They design a statistical model which uses n-gram to represent the document content and then determine the documents is benign or not by statistical-learning analysis. Some other studies focus on scanning shellcode or embedded objects in the document to detect malicious documents. Schreck *et al.* [49] presents a new approach named BISSAM to detect malicious code in the documents. They extract malicious shellcode from the Office files, and then automatically determine which type of vulnerabilities the malicious documents used. Chen *et al*. [24] designs and implement a malicious document detector called Forensor to differentiate malicious documents. They introspect file formats to retrieve objects inside the documents, and to automatically decrypt simple encryption methods to discovery potential shellcode. In addition, there are also many free forensic tools which can be used to extract the malicious code from the document. OfficeMalScanner [11] is a famous tool which can be used to scan for malicious trace. It can be used to detect external embedded data in documents which are considered suspicious, such as shellcode, PE-files or OLE streams. These methods based on binary content analysis often come to fail when some evasion techniques (e.g., polymorphism, obfuscation) are used to hide the patterns or make the detector fail to parse the document.

The third methods are built on structural feature extraction which are often combined with machined learning algorithms. These methods often extract features from structural data or metadata from the files, and then use them to train a detection classifier. Malware Slayer *et al*. [41] uses the linearize path to specific PDF elements to build maldoc classifier. Smutz *et al*. [51] presents a new approach for detection of malicious documents through machine learning. Their approaches are based on features extracted from the metadata and structure of the PDF samples. Hidost [52] is a static machine learning-based detector which combines the logical structure of files with their content for even better detection. Maiorca *et al.* [39] presents a novel machine learning system for the automatic detection of malicious PDF documents. They extract information from both the structure and the content of the PDF file. Cohen *et al*. [29] presents a novel methodology (SFEM) which is used to extract structural features from the XML-based Office documents. It is the first methodology used to extract discriminative features from XML-base Office. A new framework named ALDOCX was also implemented based on their methodology [45].

The feature extraction methods combined with machine learning are more robust. However, they also suffer from some drawbacks. For example, those methods heavily rely on the structure or meta-data of the file. Therefore, they can only be used to detect some specific formats which are easily extracting features from the structural data, such as PDF, DOCX. What's more, these methods are strongly influenced by the structure or metadata of the document. They can be easily escaped by adversarial attacking [31; 57]. Attackers can modify an existing malicious document and make it appear more benign by just adding empty structure or metadata items into the document.

## 2.2. Dynamic-based analysis

Other important methods for detecting malicious documents are based on dynamic detection. The principle of dynamic technique is to open the document in a sandbox environment and analyzes the behavior of the target process to decide whether it is proper or not. The approaches based on dynamic analysis can be further separated into three categories.

In the former category, run-time behavior of a document viewer is analyzed. Daniel Scofield .etc. [50] employ features correspond to the runtime interactions that a document viewer to train a classifier and then use it to detect malicious document. Meng Xu *et al*. executes a document across different platforms and monitor the viewer's behavior. Then they use behavioral discrepancies as an indicator for malicious documents detection [56]. What's more, some methods monitor the APIs called by the target process, and then determine whether the behavior is normal or not by analyzing the API sequences. Those methods are widely adopted in some anti-virus software, such as FireEye and Wildfire. The shortcoming of those methods is it is very difficult to design rules to identify complex behavior [35].

The second category is based on exploring the characteristic of the exploitation process, including exploit detection by Control Flow Integrity (CFI) and ROP detection. Exploitation detection by CFI will generate a complete control flow graph (CGF) by performing static pre-analysis, and then check each control-flow transfer of the target process. If the transfer is not part of CFG, then it can be identified as potential anomaly [58]. Unfortunately, these approaches are often effective in practice due to various reasons. Firstly, constructing a complete CFI needs to access the source code, while it is not available in some times. Secondly, it is hard to check the validity of some transfers which occurred in dynamic code. Return-Oriented Programming (ROP) is a prevalent technique which is widely used to break the protection mechanism Data Execution Protection (DEP) providing by the operating systems. The attackers do not inject any new code and just use the existing sequences of instructions in the process's memory, called gadgets to chain together to perform the malicious behavior. ROP attack often has a continuous jump on those Gadgets when it is executed. Therefore, most methods are aimed at detecting gadgets execution through dynamic instrumentation or by leveraging existing hardware branch tracing features. Some researchers use rule-based solutions to detect ROP attack by identifying continuous ret instruction in the execution flow of the program. Emily R. Jacobson, *et al.* defines conformant program execution (CPE) as a set of requirements on program states. Then the code reuse attack violate these requirements can be detection [34]. Yueqiang CHENG, *et al.* designs a novel system, ROPecker, to detect ROP attack at run-time by checking the presence of a sufficient long chain of gadgets in past and future execution flow [27]. Meining Nie, *et al.* proposes a practical exploit early detection system called Xede to comprehensively detect code reuse and code injection attacks [43]. Some other researchers use Anomaly-based detection to against ROP attack. They first learn a baseline of normal behavior and then detect attacks by measuring the statistical deviation from the normal behavior [32].

The last category is based on taint analysis by employing a dynamic tracing technique to detect exploits [30; 42]. It marks the input data as tainted data, then monitor the program execution to track the data flow in the program and check whether the input data have been used in a dangerous way. However, it suffers from the high overhead. Then it cannot be accepted in real practice.

## 3. Detecting malicious documents based on entropy time series

For a better understanding of the detection of malicious documents based on its entropy signal, a brief view on the process of email cyber-attack based on malicious documents was given, as showed in Fig. 1.

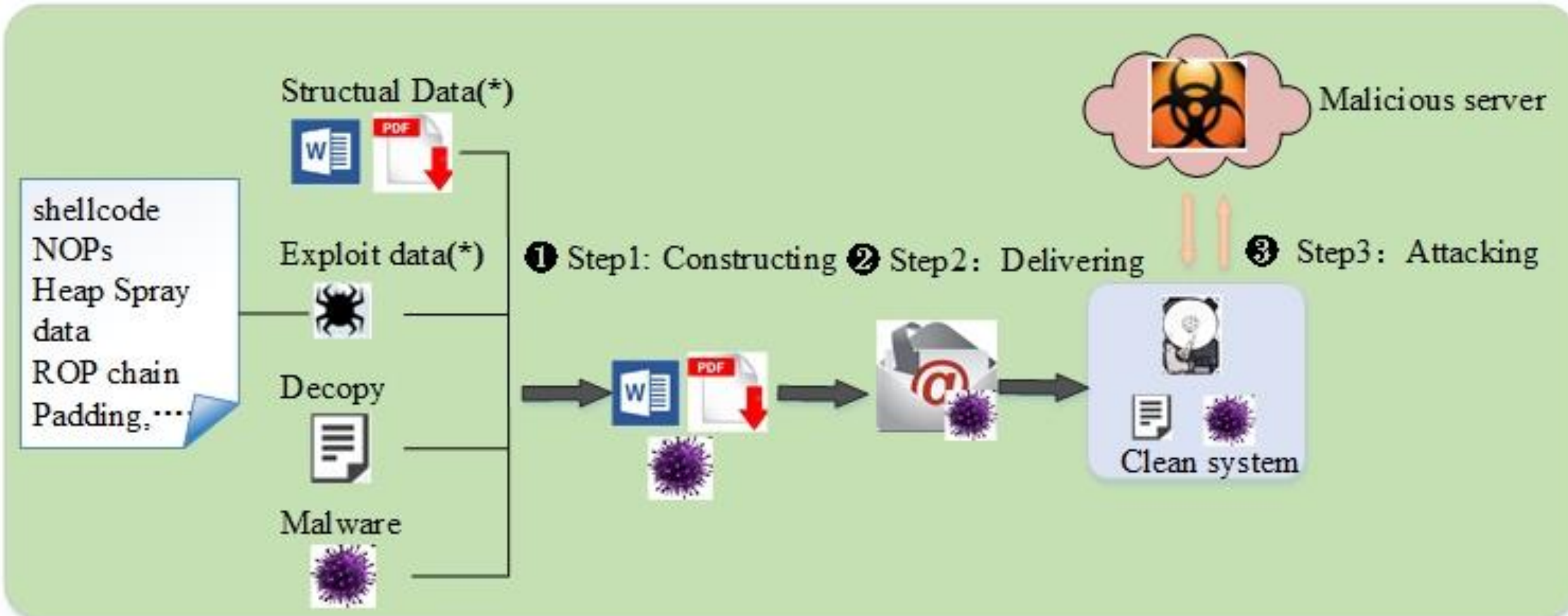

Fig. 1. The main process of a cyber-attack based on exploiting document.

The process of a cyber-attack based on malicious document consist three main steps: constructing, delivering and attacking. Firstly, Attackers construct an exploiting document aimed at triggering the vulnerability of the target software to install malware. The component for the exploiting document often come from four parts: structural data, exploit data, decopy (a benign document), malware. The structural data is critical because the malicious document must meet the requirements of the formats of the file. The embed exploit code is also indispensable which helps the attacker to launch attack. It often contains many parts, including vulnerability trigger code, ROP chain, padding data, NOPs instruction data, Heap Spray data, shellcode and so on. The trigger code is used to trigger the vulnerability of software. While padding data and Heap Spray data are used to fill in some memory spaces to guarantee the attacker can hijack programs EIP to execute gadgets and shellcode. ROP chain is a set of instruction addresses which are come from the target process, and they are combined together to construct a ROP chain. Shellcode is a piece of operation codes (Opcodes) aimed at performing specific malicious behavior (e.g. downloading or running a malware). In general, most malicious documents are embed with a benign document (deopy) and a malware. Normally, the decopy and malware will be encrypted in order to bypass the anti-virus software. When the attackers have constructed the malicious documents, they would deliver them to their targets by emails. The vulnerability would be triggered and then the shellcode would be executed to release the decopy and the malware into the disk when the document was opened. Lastly, the malware would be executed silently and the decopy would be opened automatically to deceive the victims.

In this research, we designed a framework which can be utilized to detect malicious documents based on machine learning methods. As stated early, it is hard to extract features from the structural data except for PDF and XML-Based office documents (Offic07). Also, it cannot feed a ML or DL algorithm with stream bytes directly like the image classification. Therefore, we must find a universal and adaptive approach that can extract features from the documents. As described before, the foremost difference between benign and malicious documents is that malicious documents are embedded with external malicious code while benign ones just contain vital structural and content data. What's more, malicious ones have less content data compared with benign documents. In fact, those malicious code is often arranged together and then embed into the different position of the target documents according to the vulnerability they exploit. There are plenty of similar features among the embedded external data. For example, the shellcodes often contain the same byte sequences and they often perform the similar functions (e.g. downloading malware from host, releasing the embedded malware on the disk and executing it). Gadgets addresses used to construct the ROP chain often comes from the same library. What's more, many authors often encrypt or obfuscate the malicious code. But simple encrypt algorithms (e.g. XOR) are chosen in order to guarantee the efficiency. Therefore, we hypothesis that embed data will change the entropy distribution of the document, which makes them different from the clean data. Then we decide to detect the exploit document based on the entropy distribution, which we called structural entropy.

The concept of structural entropy often can be seen in other fields of computer security, especially for packer detection and malware detection [19-21; 23; 46; 54; 55]. In the research of packer detection, if a PE file is packed, some sections of the file will be encrypted or compressed. Then the mean entropy of the file will be higher than the unpacked file. While, the features extracted from the file merely comprise mean entropy of the file or a few subcomponents just can distinguish a PE file is unpacked or not and is not adequate for identifying a malware. Researchers found that many malware authors will share some similar techniques (e.g., encryption or compression of the row bytes, insertion of NOP operators and frivolous instructions, randomization of instructions) to make the malware can bypass the anti-virus and these behavior will change the entropy signal of the file. They also found that those patterns maybe characterized not only by max entropy value, mean entropy value, etc., but also by their homogeneity. Then researchers start to utilize structural entropy to measure the similarity among the malware. They calculate the structural entropy of malware and represent it with entropy sequence. Then some comparison algorithms based on entropy sequence can be used to assess the similarity between malicious and clean files. For example, Baysa *et al.* [21] uses wavelet analysis on the entropy sequence and then measured the similarity among malwares by using sequence alignment. Canfora *et al.* [23] exploits the structural entropy to detect Android malware. They represent the DEX file with entropy sequences and then compare the similarity score based on the Levenshtein distance. In addition, some researchers find that the file's entropy sequence can be treat as a signal then many signal processing methods can be applied to extract more information to help for identifying malicious files. Patri et al. [46] uses the entropy time series to represent the content of the file and applied the shapelets to identify malware. Wojnowicz *et al.* [54] sees the entropy sequence as a non-stationary time series and identify suspicious patterns of entropy change through many ways. They extract features from the entropy signal based on Wavelet Energy Decomposition, Detruded Fluctuation Analysis and Mean Change Point Modeling. Then these features are combined with some string features to enhance the ability for detecting malwares.

There are plenty of similarities between malware and malicious documents detection. In fact, the formats of Office file or Acrobat are highly structured. A benign file should contain structural data and content data, while malicious documents just contain the structural data and malicious code. So the entropy distribution of the malicious document may be significant different from benign files owing to the embedded malicious code. Figure 2 shows some examples of the entropy distribution of clean and malicious files.

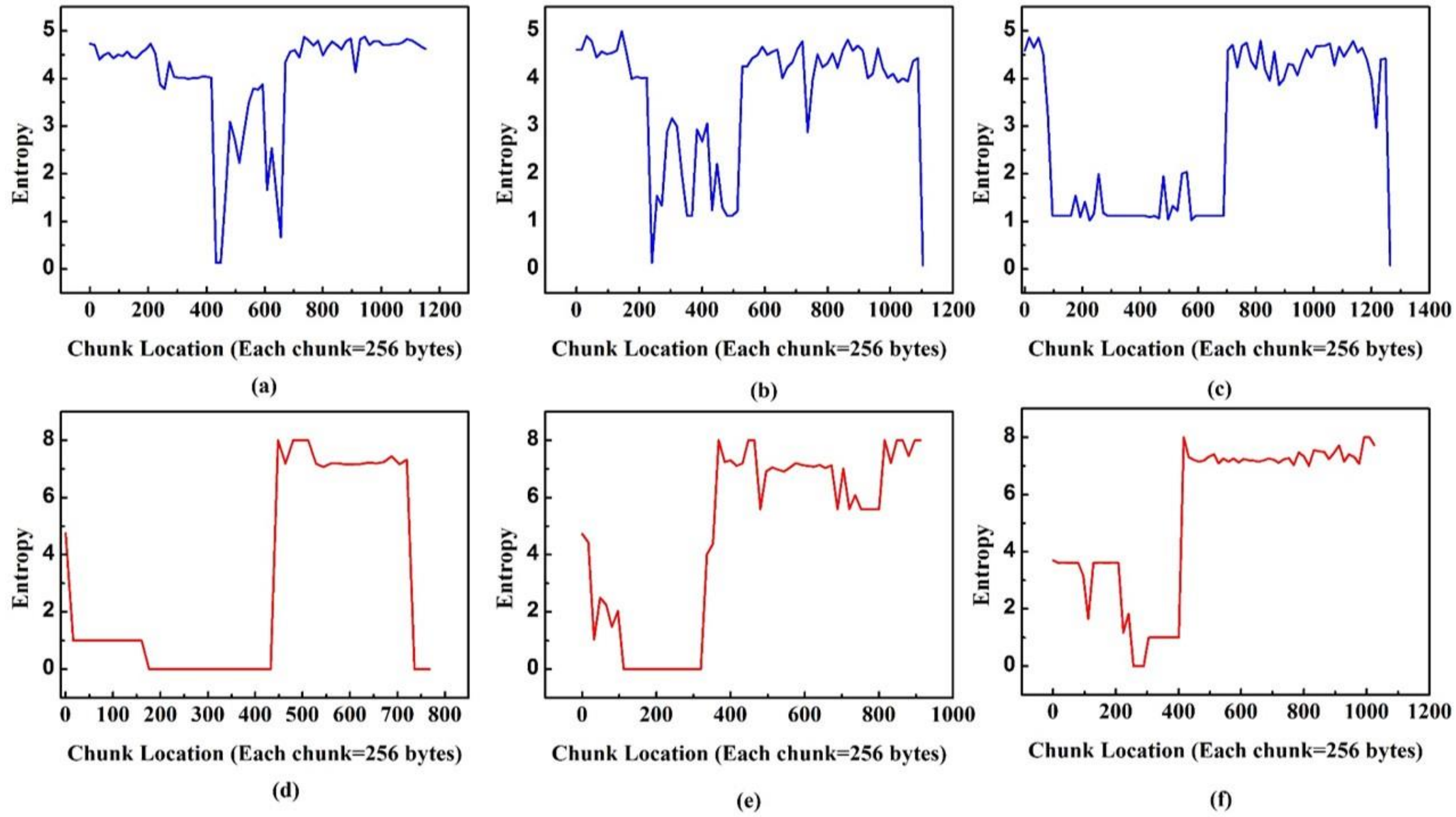

Fig. 2. A sample of the entropy distribution of benign documents and malicious documents. The subgraph (a-c) on the top are the entropy distribution of three clean documents. While the subgraph (d-f) on the bottom correspond to malicious documents.

The subgraph (a-c) represent the entropy time sequence of benign files, while the subgraph (d-f) describe the entropy sequence of malicious documents. From the perspective of visual observation, it can clearly see that there are some significant differences on the entropy distribution between benign and malicious files. While our goal is to identify the characteristics or patterns hidden in the entropy distribution that can be used to effectively distinguish benign and malicious documents through mathematical statistical analysis. In this research, we treat the entropy as a time series, which called ETS (entropy time series) like many researches on malware detection. Then we extracted features from the signal. Specifically, we extracted both global and structural entropy features. For global features, we extracted some statistical characteristics, such as Length, Mean, Max-value, Stdev, Max-percentage. For structural entropy features, we translate the raw time series into a vector representation by two methods: discrete wavelet decomposition and BOW. Features extracted by DWT show the detail characteristics about the entropy time series' changes. And the methods based on BOW can extract the local features from the entropy sequence. As describe before, the malicious code is inserted into different position, and most of them share the same characteristics. There have a lot of similar local features on their entropy distribution. Therefore, we extract those local features by using BOW which make the classifier performs better. The main framework is illustrated in Fig. 3, which contains two steps: training and testing.

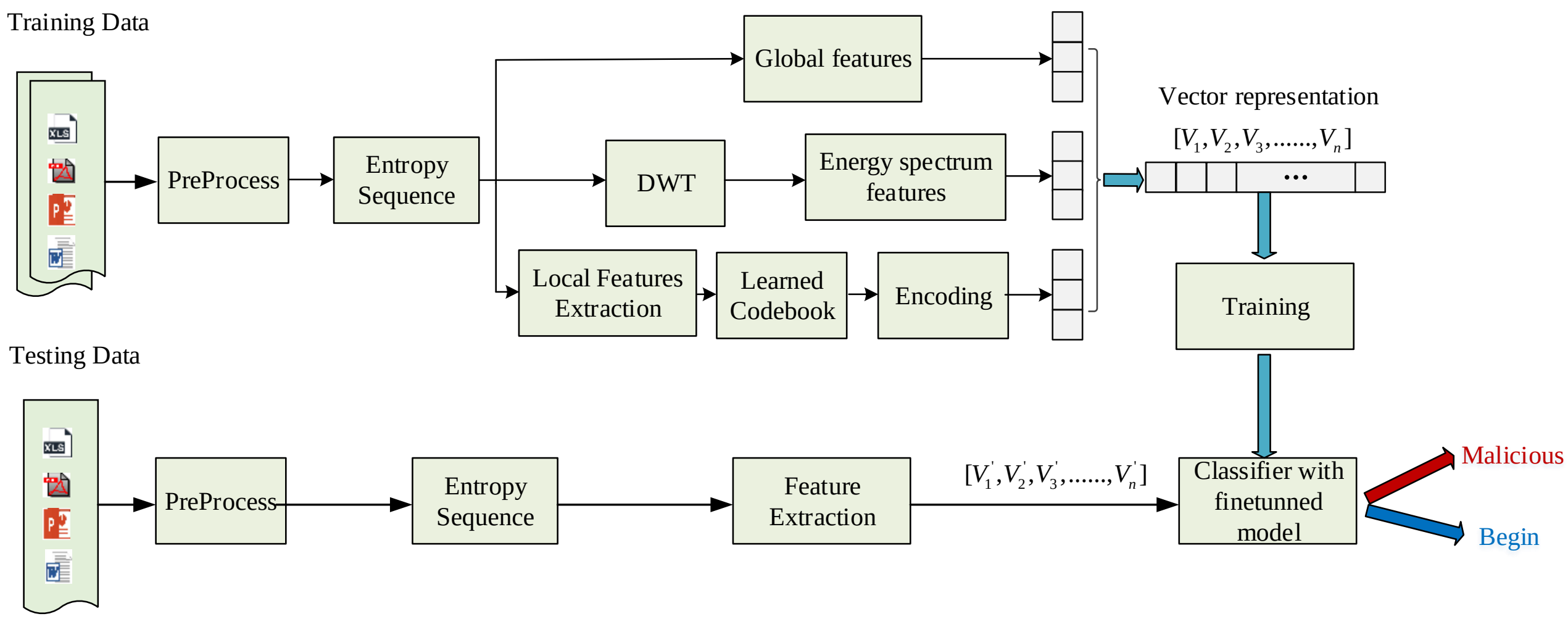

Fig. 3. The main framework of our method. It mainly includes two processes of training and testing.

In the training step, we first collect numerous samples working as our training data. A preprocess is applied before we extracted features from the documents which used to decompress some Office07 files, such as .dox, .xls, or some compressed objects in PDF documents. Then we calculate the entropy sequence and get the time series. Next, we extract features from the time series and represent them with feature vectors. In the end, a classifier is trained based on those feature vectors. For the testing step, features are extracted from the new documents just as the same way in the training stage. Then we fed them into the fine-tuned classifier to identify its properties.

# 4. Methods

## 4.1 Entropy time series of the document

In order to calculate the structural entropy of the file, we first divide the file into size fixed blocks, and then calculate entropy value of each blocks. Specifically, the steps are as follows:

- We slide a file with a non-overlapping window and divide the file into many blocks.
- We calculate the entropy value of each window.
- We array the entropy value based on the location of the blocks and get the entropy sequences.

We determine entropy sequences which denote $V = v_{1,} v_2, ......, v_N$ , where $N$ is the number of windows and $v_i$ is the entropy value of every block. In practice, we use a window with the length of 256. If the last bock is not aligned with 256, we deal it with the following rules: we padding it with zero if the length of the last block larger than 128, otherwise, we discard the last block. In each window, we calculate the entropy value like this:

$$H(X) = -\sum_{i=1}^{255} p(x_i) * \log_2 p(x_i) \qquad (1)$$

The range of the value of $v_i$ is 0 to 8. When every bytes in the window are equal, then $v_i$ is 0. If all bytes are different in the window, the value $v_i$ will be equal to 8.

## 4.2 Features extraction based on the discrete wavelet decomposition

Discrete wavelet transform (DWT) is commonly used in image feature extraction, and non-stationary time series signal processing and so on. Compared with Fourier transform, the strategic advantage of DWT is that it can capture both frequency and location information, by examining the signal at different scales. DWT is able to extract "detail" exhibited in the signal. By using the output of the DWT (the so-called "wavelet coefficients"), it is possible to obtain a series of coarse-to-fine approximations of the original function.

In the wavelet transform, there are two functions that play a primary role in wavelet analysis: the wavelet function (mother wavelet) and the scaling function (father wavelet). In the process of decomposition, a signal will be represented through a linear combination of their wavelet functions and scaling functions. In this study, we apply Haar wavelet's wavelet function and scaling function to decompose the signal. Haar wavelet is the simple family of wavelets where the wavelet functions $\psi_{HAAR}(t)$ and scaling function $\phi_{HAAR}(t)$ can be described as follows:

$$\psi_{HAAR}(t) = \begin{cases} 1, 0 \le t < \frac{1}{2} \\ -1, \frac{1}{2} \le t < 1 \\ 0, t < 0, t > 1 \end{cases} \qquad (2)$$

$$\phi_{HAAR}(t) = \begin{cases} 1, 0 \le t < 1 \\ 0, otherwise \end{cases} \qquad (3)$$

Before decomposing the signal, it needs to produce a set of shifting functions of the wavelet function and the scaling function. For this process, there are different translations and dilations of the mother and father function. Given the mother function $\psi(t)$ and father wavelet function $\phi(t)$ , we can construct shifting functions from the wavelet function and the scaling function with two parameters: scaling and translating. Formally, the set functions of wavelet function and scaling function can be defined as follows:

$$\psi_{j,k}(t) = 2^{j/2}\psi(2^j t - k) \tag{4}$$

$$\phi_{j,k}(t) = 2^{j/2}\phi(2^j t - k) \tag{5}$$

Here, $j$ is the parameter of the dilation which denotes the level of resolution at a particular stage, and $k$ is the parameter of the position.

During the Wavelet decomposition, an original time series is decomposed into a sets of coefficients which represent the detail and of the signal. For a given signal $x(t)$ who has $T$ points $(t = 1, ..., T)$. We rescale the signal to ensure that the signals start from 0 to the end of 1. Then we can calculate the detail coefficients (also named mother wavelet coefficient) and approximate coefficients. The detail coefficients is generated by the inner product of the signal and the wavelet function. The formulas are presented in formulas 6.

$$d_{j,k} = \left\langle x, \psi_{j,k} \right\rangle = \sum_{t=1}^{T} x(t)\psi_{j,k}(t) \tag{6}$$

While the approximate coefficients is produced by the inner product of the signal and the scaling function. It can be described as follows:

$$a_{j,k} = \left\langle x, \phi_{j,k} \right\rangle = \sum_{t=1}^{T} x(t)\phi_{j,k}(t) \tag{7}$$

In fact, approximate coefficients represents the approximation of the signal and the mother wavelet coefficients store the detail of the signal. DWT allows for Multi-resolution Analysis of the signal. The coefficients at different levels provide a description from a coarse approximation to fine-grained approximation. For each level, the previous approximate coefficients are further decomposed into new coefficients. Specifically, the signal $x(t)$ can be decomposed into a series of approximations $x_j(t)$. The process of decomposition is a recursive, where by each level, the approximation $x_{j+1}(t)$ is a more detailed refinement of the previous approximations $x_j(t)$. We can get the functional approximations through the wavelet coefficients as follows:

$$x_{j+1}(t) = x_j(t) + \sum_{k=0}^{2^j - 1} d_{j,k}\psi_{j,k}(t) \tag{8}$$

Using the discrete wavelet coefficients as features to classify a time series has been used in other fields. In those fields, it is ideal to use the coefficients as the input to the machine learning model. However in our case, we cannot do like this. As the length of the documents are different, then the coefficients can also be varied from one to ones. However, in machine learning mode, the input feature vectors must be having a fixed length. In order to address this issue, we use the Wavelet Energy Spectrum as the features to represent the time series. The Wavelet Energy Spectrum reflects the energy distribution at different resolution levels and it reveal the "detail" or "variation" available at various resolution levels. Then energy spectrum is computed as a function of the wavelet coefficients $d_{j,k}$. In particular, the "energy", $E_j$ for the time series at the $j$ th resolution level is defined by:

$$E_j = \sum_{k=1}^{2^j-1} (d_{j,k})^2 \quad (9)$$

In DWT with Haar wavelet, the parameter $j$ is required to be dyadic length (power of 2). This parameter provides $j$ multi-scale analysis for different positions of the entropy time series. For example, if we want to analyze a file from scale 1 to 5, then parameter $j$ can be settled from 2 to 32. Similarly, the input time series must also be a dyadic length vector. Figure. 4 shows the example of two files which we decompose the entropy sequences from level 1 to 5.

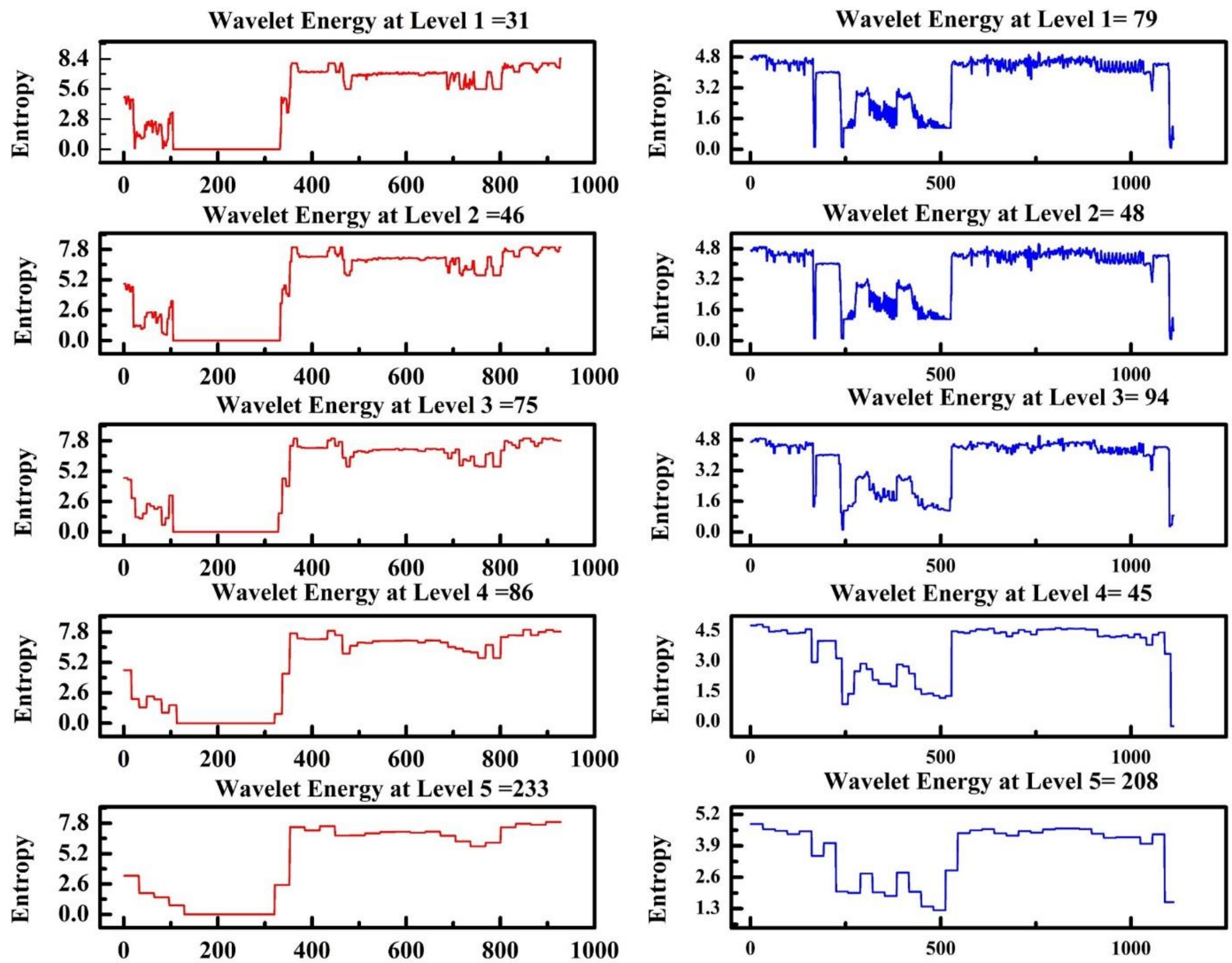

Fig. 4. The example of two files where the entropy sequences are decomposed from level 1 to 5. The left side is the enrgpy spectrum of malicious document, and the right side is for benign one.

From Fig. 4, we can see that the wavelet energy for malicious has great changes in high level. While for benign files, the great changes happened in both coarse and fine-grained level.

In this study, a sample which has $T$ points in its entropy sequence will be extracted $J = [\log_2 T]$ features in its wavelet energy spectrum. Then the dataset will be split into different groups and each group has the same number of features. For example, a group files with the length $2j$ to $2j+1$ will have the $J$ features. While different group will have different numbers of features. In order to make all the samples have a fixed length features. We set the $J_{\max} = 20$ which means we create 20 features (means the maximum file length will be $2^{20}*256=1048576$, about 20M) from the time series. In fact, the length of most simples in the dataset is less than 20M. If a sample with less than 20 features, we will pad it with zero. Finally, we total extract 20 features from the file entropy sequence based on DWT.

## 4.3 Features extraction based on the BOW

BOW is a widespread technique in text mining for document representation. In this model, a document will be presented as the bag of it words (we often called it codewords). The BOW model is only concerned with whether known words occur in the document, not where in the document. In this process, the order information will be ignored, while the multiplicity will be kept. In the past, BOW has been extended to analysis of images and videos [33; 44]. In the task of exploiting document classification, samples may have different structures as they may exploit different vulnerabilities. While they may share many small components and most of those data share the similarity personalities. Therefore, there would be hide many similar local patterns in the entropy time series among the malicious documents. As the similar local patterns will be embed in different position into the file. Therefore, the order information is not very significant for us. Then we decide to use BOW to extract the local patterns hiding in the time series.

A typical BOW framework consists of three major steps :( 1) Local feature extraction, (2) Codebook generation, (3) Histogram computation. Figure. 5 illustrates the main procedure of our method:

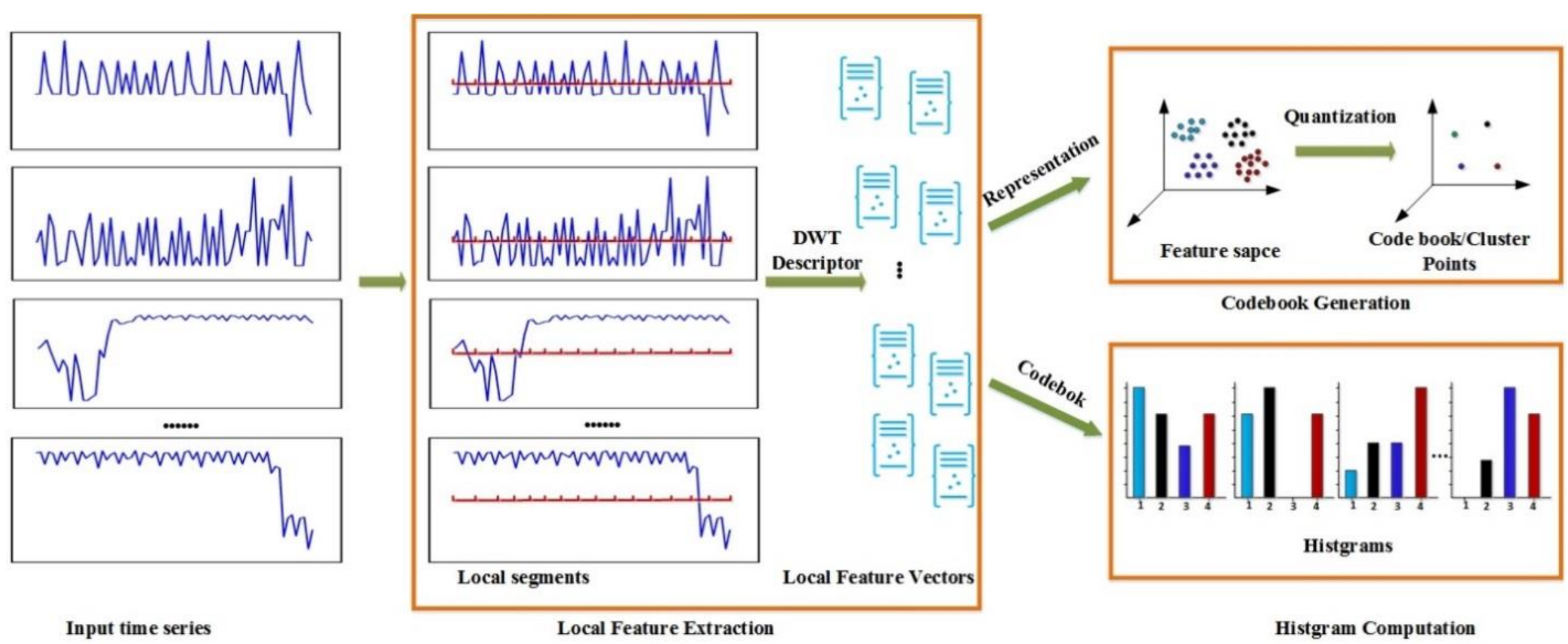

Fig. 5. The main procedure of time series vectorization based on BOW.

**1. Local features extraction.** In the first step, we split all the time series into a group of local segments where each segment has the same length, and then we extract features from each segment. Regarding this procedure, successful feature extractors have been developed, like SIFT [38], Space Time Interest Points (STIPs) [36]. Nonetheless, those methods are invented to process images. For time series classification, Wang. Etc. [53] use DWT to describe each local segment. In this study, we use the DWT to decompose each segment and then use the approximate coefficients as features vectors to represent the local segment. In the previous section, we know that the approximation wavelet coefficients represent the approximation of the time series. Therefore, it can be used to represent the local features for each segment. The detail wavelet coefficients have been used to represent the detail information of the file in subsection 4.2. Then the approximation wavelet coefficients can be used as a better supplement to extract local features. In this step, we also use the Harr wave function to decompress the local segments. The decomposition level is based on the length of the segment. In our research, we finally set the segment length with 6. The coefficients getting at different levels are combined together and then used to represent the local segment.

**2. Codebook Generation.** In this step, we generate a predefined a codebook which contains all the code words occured in the training data. The BOW counts each code word which exists in each document and then uses the histogram of code words to represent each time series. In general, clustering algorithms are applied to generate clustering centers on the local features, which regarded as base element of the codebook. In this research, we use the K-Means to cluster all the local features. The K-Means algorithm divides the training dataset into disjoint clusters $C_k$ where each cluster is characterized by the mean value of the sample $U_k$. These mean values often referred to as clustering centers. For example, we have a group of local segments $X = [x_1, x_2, ..., x_n]$, where $x_i \in R^d$, are local features extracted from training data, the construction of codebook can be modeled as an optimization problem, and the objective function is formulated as follows:

$$J(c^1, c^2, ..., c^m, \mu_1, ..., \mu_k) = \frac{1}{m}\sum_{1}^{m}\left\| x^{(i)} - \mu_c(i) \right\| \quad (10)$$

Where $\mu_i$ is the clustering center. The codebook $\mu$ has $k$ codewords, and each code word is a d-dimension vector, which has the same length as the local features.

If we use all the local features extracted from training data to feed the K-Means algorithm, the cost of computing will be intolerable. Therefore, in our research we just select some local features to cluster. Specifically, we just randomly select 20% of the local features to cluster.

**3. Histogram computation.** This step aims to encode a sample with the generated codebook and then represent it with its histogram. A time series will be first divided into many segments, then each local segment will be assigned with a code word which has the minimum distance with the local segment. Specifically, if the codebooks contain $k$ codewords, where each words is $d$ dimension, $\mu = [\mu_1, \mu_2, ..., \mu_k]$. A local segment with local features $x_i$ is assigned the $c$ th codeword that: $c^* = \arg\min d(\mu_j, x_i)$), where $d$ is the Euclidean distance between $\mu_j$ and $x_i$ . Each time series can be represented as a histogram of codewords. Lastly, we normalize the histogram of the codewords when all the local features have been assigned codeword. Then it can be utilized to represent the local feature of the time series. Figure. 6 illustrates the example of the BOW representation for some malicious documents.

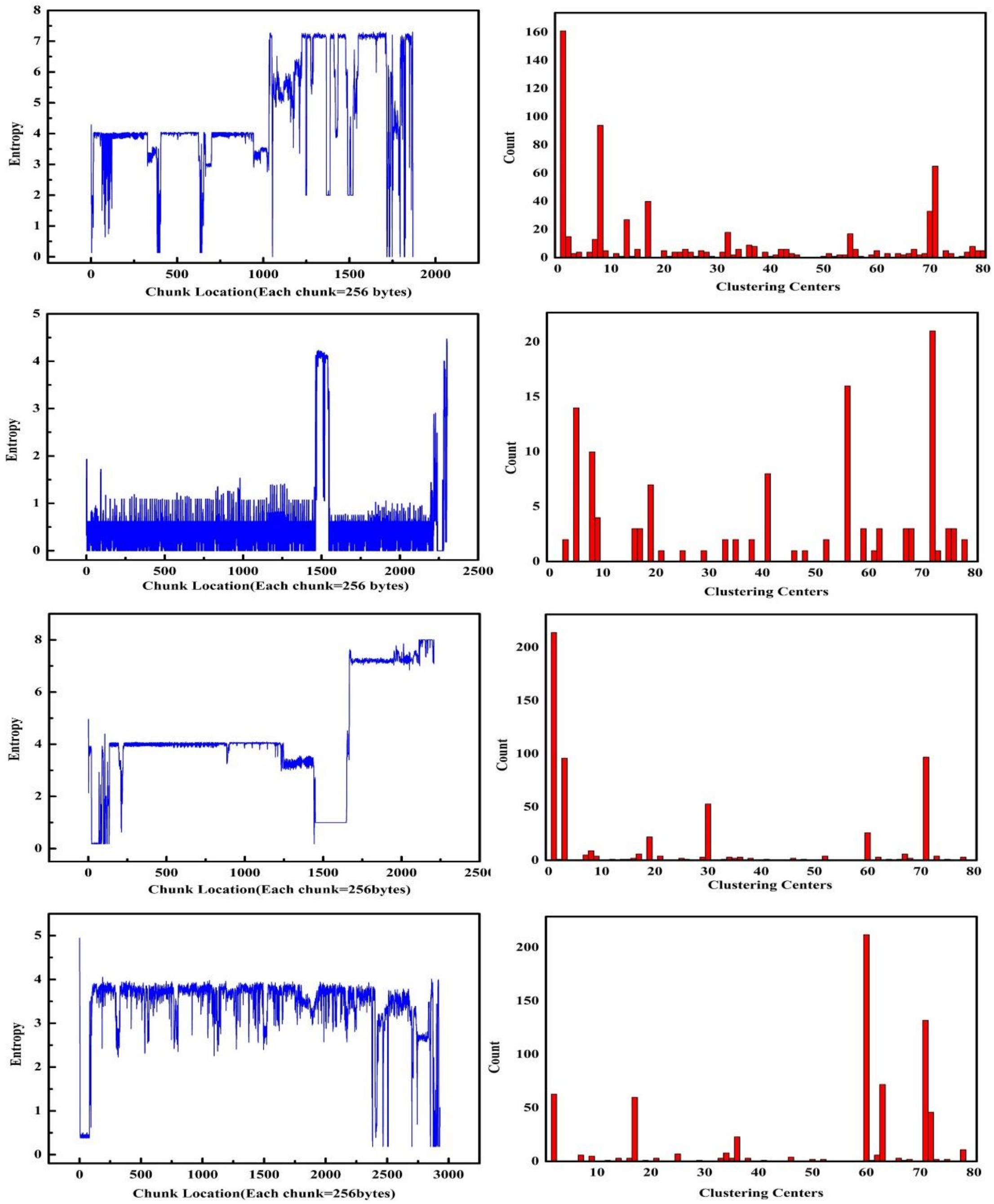

Fig. 6. A sample of the BOW representation for malicious documents. The left images are the entropy distribution of the documents and the right images is the corresponding histogram of codewords.

## 4.4 Global Features

We also extract some global features from the raw time series, like the method using the entropy signal to measure malwares [54]. The extracted features can provide global information about the time series which can enhance the ability of the classifier to detect malicious documents. The features we extracted are as follows:

- Length: Length of the entropy time series.
- Mean: Mean value of the entropy time series.
- Stdev: Standard deviation of the entropy time series.
- Max-Value: Max value of the time series.
- Max-Percentage: Percentage of the time series who has a value larger than 7.0.
- Zero-Percentage: Percentage of time series who has zero value.

# 5 Experiments and Evaluate

In order to evaluate whether the entropy distribution of the file has a good contribution to the decision of malicious documents. This section we will describe our experiments and analyze the results. We have conducted many groups of experiments. Our experiments are centered at answering the following three questions (RQs).

- RQ1: Does the extracted features can effectively reflect the nature of the malicious documents? And which machine learning algorithm can be more effective for the classification tasks and what is the best configuration for the representation of time series and the classifier? For answering this question, we conduct cross-validation experiments on a created dataset. We extracted features from the training data and then apply different classifiers to assess their performances (subsection 5.4). For the configuration of the time series representation, we do experiments to select the best size for two parameters: local segment and codebook (subsection 5.5). For the selected classifier, we also do some experiments to select the best configuration (subsection 5.6).
- RQ2: Is the combination of the three type’s features more conducive to the classify task? For answering this question, we analyze the contribution of the three features and their combinations (subsection 5.7).
- RQ3: How is effective is our method when compared with other approaches or tools? For answering this question, we compare our method with other 15 leading anti-virus software and some prevalent tools (subsection 5.8).

The environment for all the experiments is as follows:

- Hardware: Intel(R) CPU E3-1231, 3.40GHz, 8.0GB.
- Software: Operating system (Windows7 SP1 (64-bit)); Python (version 2.7.13); sklearn (version 0.19.0); xgboost (version 0.7.2); pywt (version 0.5.2, a wavelet transforms module in python).

## 5.1 Dataset

To better evaluate our framework and methods, we created a large dataset which contain Microsoft Office and PDF files. We acquired a total of 10381 samples, including 5344 malicious files and 5037 Benign files. Those samples collected from different sources are presented in Table 1.

Table 1. The dataset from different sources

| Collection Source | Year | Malicious files | Benign files |
|---|---|---|---|
| Virus Total | 2014-2015 | 3740 | 0 |
| contagion | 2013 | 890 | 0 |
| Tracker h3x | 2016-2017 | 714 | 0 |
| Baidu | N/A | 0 | 1739 |
| Google | N/A | 0 | 1751 |
| Bing | N/A | 0 | 1547 |
| Total | N/A | 5344 | 5037 |

An important source for those samples is from a public dataset which was established by John et al [22]. They collected the suspicious files which were uploaded to VirusTotal between April 2014 and March 2015. The dataset contains Office and PDF documents. Those documents were uploaded by 255926 users which are from 201 countries. Because the uploaded files are not all exploiting documents. Then they have filtered out unrelated documents (e.g. macro infected or safety documents). Finally, the published dataset contains 3740 malicious documents. Parting from these samples, we also collect samples from some web sites. These sites are divided into two parts, one of which collects samples from some APT attacks and allow users to download for further research [14]. Another part are some malware repositories that offer a free service which allows people to upload suspicious files for detection and download samples which are uploaded by others [7; 8; 15]. We collected more than two thousand samples from those websites. In order to ensure all the downloaded samples are malicious, our team had checked all of them by using VirusTotal's service (A sample will be labeled as malicious when it is identified by any one anti-virus). What's more, we also collected clean documents by using internet search engines, like Baidu, Google and Bing (we used the command filetype: RTF/PPT/PDF/XLS/DOC/DOCX, etc.). Similarly, all the clean files were assured to be labelled correctly as benign by using VirusTotal service.

The dataset contains four types of files, including PDF, RTF, Office03, and Office07. Office03 is the file format which is used by Microsoft Office 2003, including DOC, XLS and PPT. While Office07 is the file format which is first been used by Microsoft Office07, including DOCX, XLSX, PPSX. Figure. 7 shows the distribution of file types.

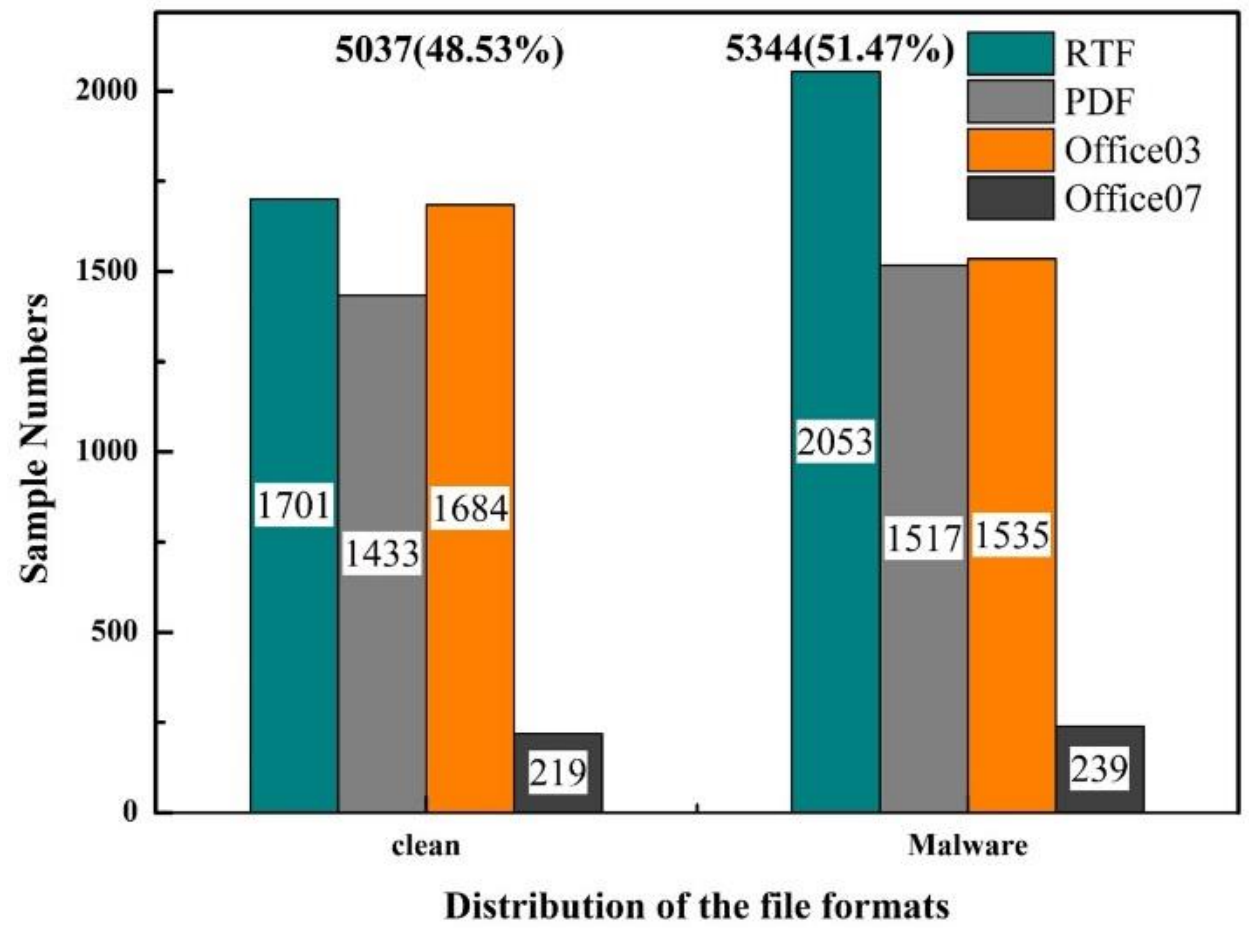

Fig. 7. The distribution of file formats on malware and clean samples.

## 5.2 Metrics for evaluation

In order to evaluate the effectiveness of our framework, we use the widely used metrics false positive rate ( $FPR$ ), false negative rate ( $FNR$ ), true positive rate or recall ( $TPR$ ), precision ( $PRE$ ), and F1score ( $F1score$ ) to evaluate it. Let **TP** represents the numbers of samples correctly classified as malicious and **FN** denotes the number of files incorrectly classified as benign. **TN** stands the number of files correctly classified as benign and **TP** means the number of files misclassified as malicious. The defined of those metrics are as follows:

$$FPR = FP/(FP + TN) \tag{11}$$

$$FNR = FN/(TP + FN) \tag{12}$$

$$TPR = TP/(TP + FN) \tag{13}$$

$$PRE = TP/(TP + FP) \tag{14}$$

False positive rate measures the ratio of false positive samples to the entire numbers of samples which are not malicious. False negative rate measures the ratio of false negative samples to the entire numbers of samples which are malicious. True positive rate measures the ratio of true positive samples to the entire numbers of samples that are malicious. Precision measures the correctness of the detect malicious documents. F1score is determined using the Eq.15, it takes consideration of both precision and true positive.

$$F1score = (2 * \Pr ecision * \operatorname{Re} call)/(\Pr ecision + \operatorname{Re} call) \tag{15}$$

## 5.3 Cross-Validation Experiment

The first experiment we conducted is a 3-fold cross-validation experiments. In this experiment, we evaluate whether the features we extracted from the entropy time series are effective to distinguish between malicious or benign documents and which ML algorithms are more suitable for this classification task.

After acquired features from the dataset, we split them into two parts: training data and testing data. The representative vectors of the training data are feed into different kinds of learning algorithms. By processing those vectors, the learning algorithm generated a classifier. At the testing phase, the generated classifier is used to detect the testing data. In the cross-validation experiments, we use 5 different Classification algorithms on our dataset to evaluate which one is more suitable for this task, which are Decision (DT), Artificial Neural Networks (ANN), Naive Bayes (NB), Logistic Regression and Support Vector Machine, these ML algorithms we used are provided by two python packages named sklearn and XGBoost. We conduct three separate cross-validation experiments, where for each experiment we randomly split our dataset into two parts: training (accounts for 70%) and testing (accounts for 30%). In each group, we randomly selected 30%, 50%, 70%, and 100% of the training data to train the classifier. Five learning algorithms are performed on those data separately. The generated classifiers are evaluated on the testing data. For each algorithm, we use the default parameters. For the representation of the time series with BOW, we set the local segment size with 8 and code book size equal to 140. When we have complete three separate experiments, the average value of each metric is the final results. The final result of the experiments is illustrated in Fig. 8. Table. 2 gives the detail result when all the training data are used to train the model.

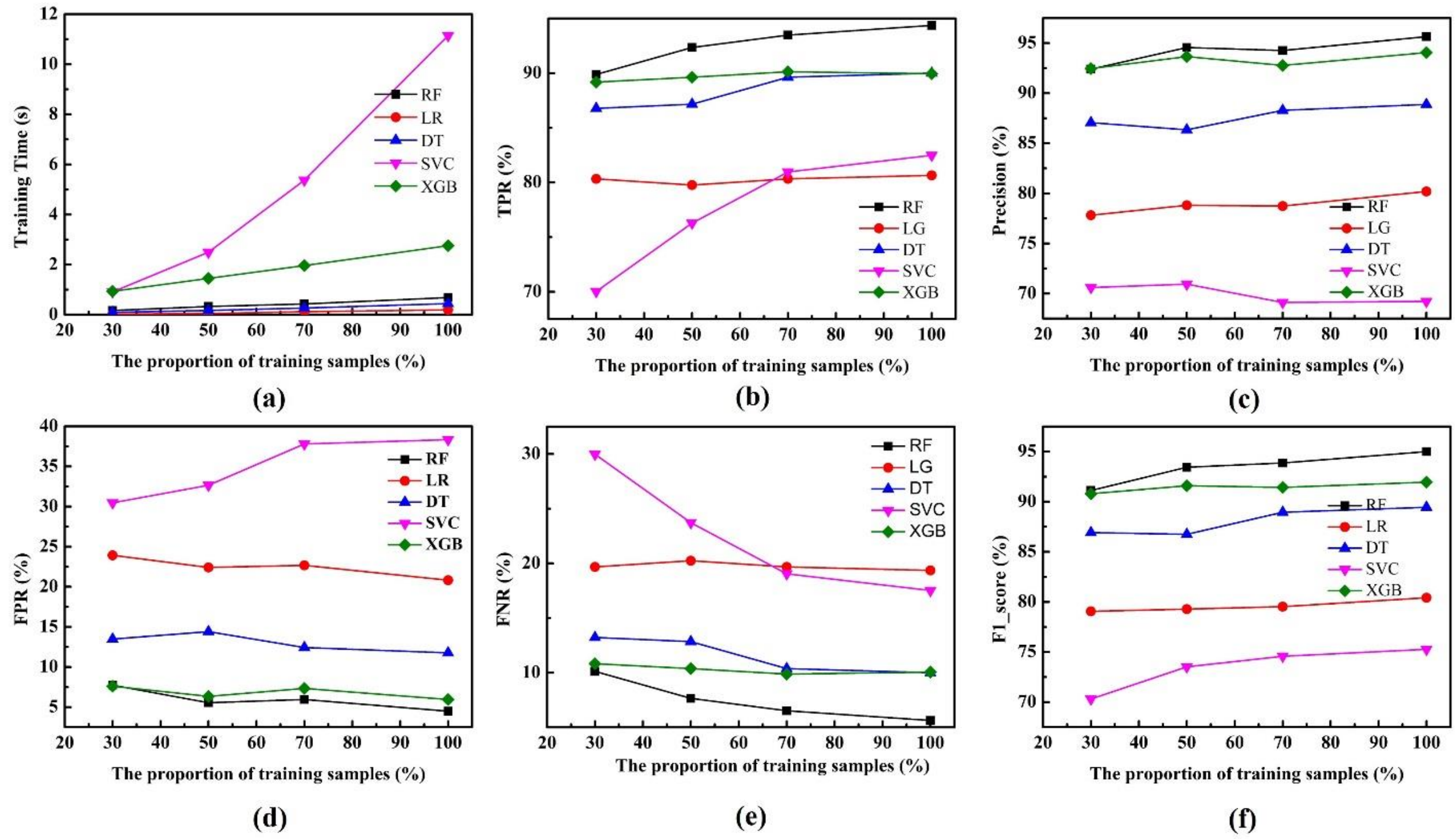

Fig. 8. The performance on five different classifiers, which are XG-Boost (XGD), Logistic Regress (LR), Decision Tree (DT), SVM (SVC), Random Forest (RF).

From Fig. 8 and Table. 2, we know two facts that the features can effectively reflect the pattern hidden in the entropy sequence and Random Forest is more suitable for this task. Although the classifier based on SVC has a poor performance, the TPR and PRE obtained are more than 70%. Overall, Random Forest classifier spend less time on training and can achieve a higher score than other classifier. With the increase of training samples, the precision and recall are increasing gradually. When all training samples are used for training the model, the generated classifier can achieve precision with 96.04% and recall with 95.12%. Therefore, in the later experiments, we use the random forest as a classifier.

Table. 2. The detail of the performance on five different classifiers when all the training data were used to train the model

| Proportion | Classifier | Training Time (s) | TPR (%) | PRE (%) | FPR (%) | FNR (%) | F1 (%) |
|---|---|---|---|---|---|---|---|
| 100% | XGB | 1.05 | 89.87 | 93.48 | 6.54 | 10.12 | 91.64 |
| | DT | 0.42 | 90.06 | 89.22 | 11.36 | 9.93 | 89.64 |
| | LR | 0.20 | 80.64 | 80.19 | 20.82 | 19.35 | 80.41 |
| | SVC | 11.34 | 82.48 | 69.21 | 38.33 | 17.52 | 75.26 |
| | **RF** | **0.61** | **95.12** | **95.66** | **4.09** | **4.84** | **95.39** |

## 5.4 Length of Local Segment and Codebook

The length of local Segment and codebook is two key parameters which are very import for time series representation. In this subsection, we performed some experiments to choose the best size of the local segment and codebook. The length of local segment must be mainly relied on the dataset. If the length is too small, it increases the cost of the computing. What's more, it also extracts some unrelated features. If the length is too large, it cannot extract features for those samples where the length of entropy time series is smaller than the segment size. In our experiments, we vary the length of segments between 4 to 22 with the step equals to 2. In the dataset, the length of entropy time series ranges from 25 to 32836. Therefore, this range can guarantee all the samples can be represented as vectors. The AUC value of those different local segment size is illustrated in Fig. 9(a).

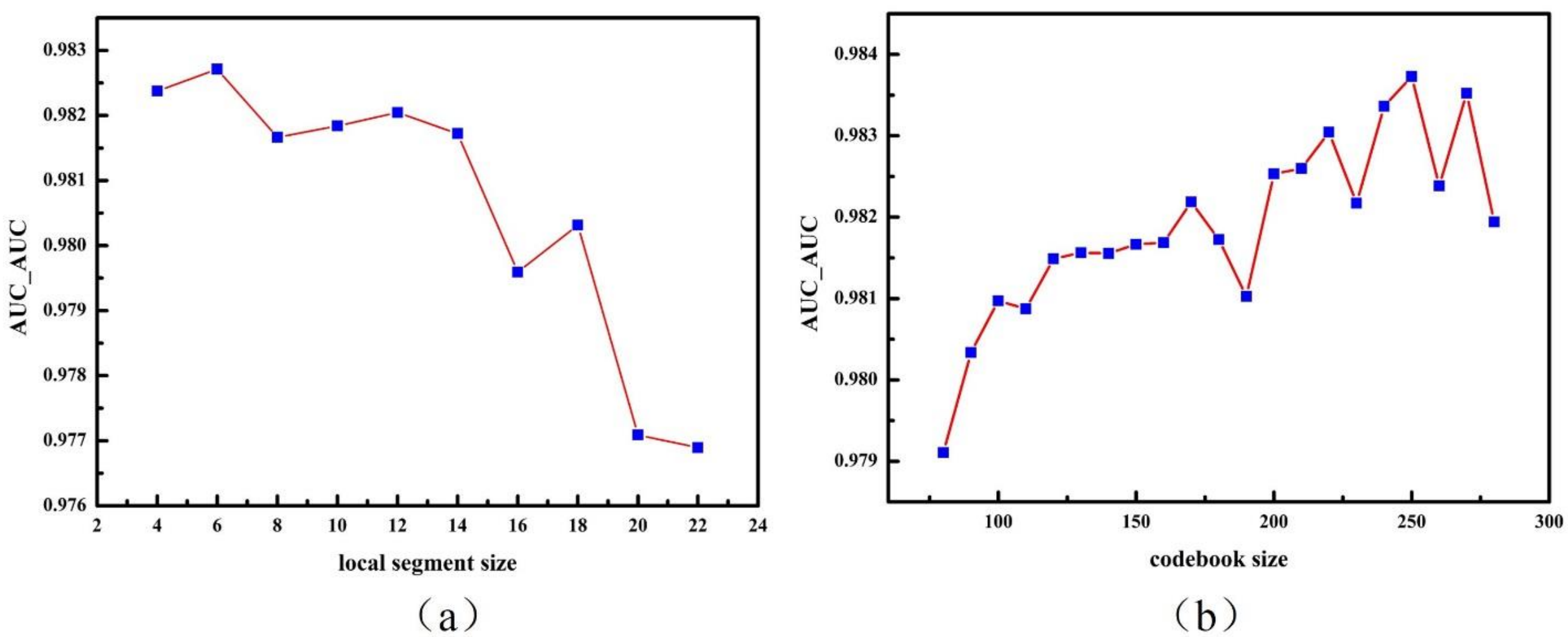

Fig. 9. A comparison of the ROC curve of different codebook size and local segment size.

From Fig. 9(a), we know that the AUC value reaches its maximum value about 0.984, when the local segment size equals 6. And then it begins to fall. Therefore, we finally set this parameter to 6 to guarantee the generated codebooks can better capture the local features. The codebooks size is also very import for the time series representation. If it is a compact codebooks, the computational cost is very low while it may lost some local features. If the codebooks is too long, it can capture more features, but the computational cost is expensive. In the experiments, we set the size of codebook ranging from 80 to 290 with step 10. The result of our experiment is described in Fig 9(b). According to the results, we know that the length between 200 to 250 is the best for the classification. If the length increases, it does not have a deeper impact on the accuracy. While the cost of computing increases. Finally, we set the codebook size with 250 for the best parameter.

## 5.5 Finding the best overall configuration of classifier

In previous experiments, we just use the default parameters of the classifier to evaluate the effectiveness of our method. This subsection, we do some experiments to choose the best parameter for the Random forest. Random Forest is an ensemble machine learning algorithm which builds multiple sub decision tress and binds them together to make the model have a more accurate and stable detection. There are so many parameters in Random forest. And all of them are either to increase the power of the model or make it easier to be trained. In this case, we tend to tune the parameters which can enhance the performance of the classifier based on the following reasons. Firstly, the classifier just need to be trained once when we constructed the model. Then it can be used to predict new documents in the future. Secondly, in this experiment, we collect about more than 10,000 samples, and each of them has 276 features (6 are from statistical information, 20 are from DWT and 250 are from BOW). Then the training time is less than 20 minutes in a complete training process where it does not contain the time we used to extract features. In this study, we choose to tune two mainly parameters which are the number of the trees (N estimator) and the maximum depth of the tree (Max depth). High number of trees gives the classifier a better performance, but the training is slower. For another parameter, Max depth, the general depth of the tree ranges from 10 to 100 according the classify task. In this study, we try many values to find the most optimum in our experiments. We use the grid search algorithm in the python package to tune the two parameters automatically. In the experiment, we use the receiver operating characteristic curve (ROC) to evaluate the performance of the classifier. For the first parameter, we set the range from 10 to 2000. In the range from 10 to 100, we use 10 as a step. For the range from 100 to 2000, the step is set as 100. The result is demonstrated in Fig. 10(a) From Fig 10(a), we know that the ROC increases significantly with the N estimator ranges from 10 to 300, while the ROC almost ceased to grow when the N estimator reaches 500. Therefore, we choose the 500 as the best parameter for the number of trees. When we fixed N estimator, we use the same method to search the best value for Max depth, we set the range from 10 to 100 with step 10. Fig. 10(b) illustrates that the ROC AUC value reaches 0.992 when the Max depth increases to 30. Then the ROC AUC value does not have a significant increase. Therefore, we finally set the Max depth to 30 as the best parameter.

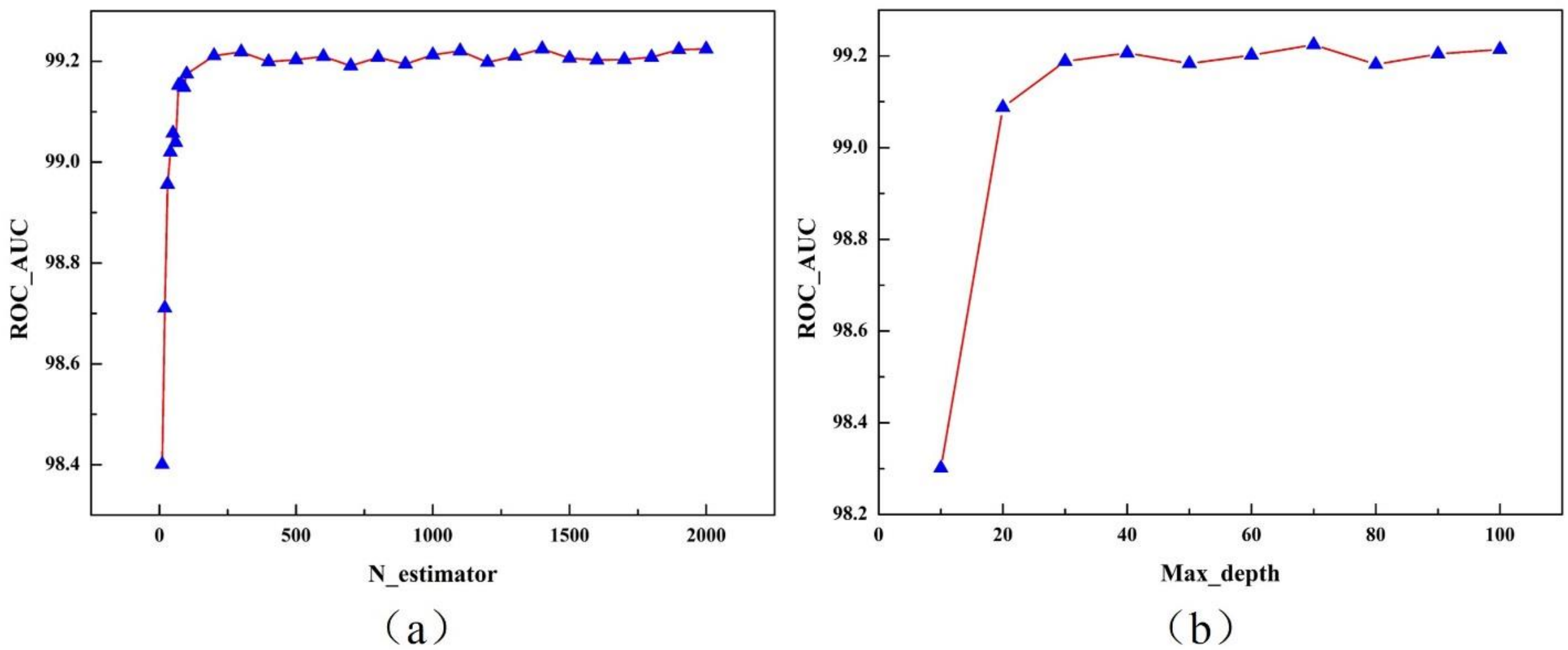

Fig. 10. A comparison of the ROC curve of different N_estimator and Max_depth.

## 5.6 The impaction of the features

In the procedure of training our model, we extract many features from the raw entropy sequence. Those features can be divided into three categories: Global features, Energy spectrum features and Local features. Is the combination of the three features more favorable for the classification task? In this subsection, we do some experiments to explore this question. Firstly, we use three types of features to train the classifier separately and the result is shown in Table3. From the result we know that, the three features have a good contribution to detect malicious document, all of them can achieve a TPR more than 80%, and the features extracted from BOW have a better performance than the other two. It can achieve a TPR with 87.90% and PRE with 93.74%. Secondly, we combined the features extracted from Global Features and DWT together, then the classifier can achieve a TPR with 87.91% and PRE 90.14%. Lastly, we add the features extracted from BOW to the feature sets, then the TPR finally reached 96.00% and the PRE reached 96.69%. In Fig. 11, we give the receiver operating characteristic curve of those experiments where we know that each individual group of features have good contribution on distinguishing malicious document, but their combination can achieve the best performance with the ROC value equal 0.992 among those experiments.

Table. 3. The performance of the model under different combination of features

| Features | TPR (%) | PRE (%) | FNR (%) | F1-score (%) |
|---|---|---|---|---|
| Global Features | 87.28 | 88.80 | 11.50 | 88.03 |
| DWT | 80.32 | 82.62 | 17.64 | 81.46 |
| BOW | 87.79 | 93.97 | 5.88 | 90.77 |
| DWT+Global | 87.91 | 90.14 | 10.04 | 89.01 |
| **DWT+Global+BOW** | **96.00** | **96.69** | **3.30** | **96.34** |

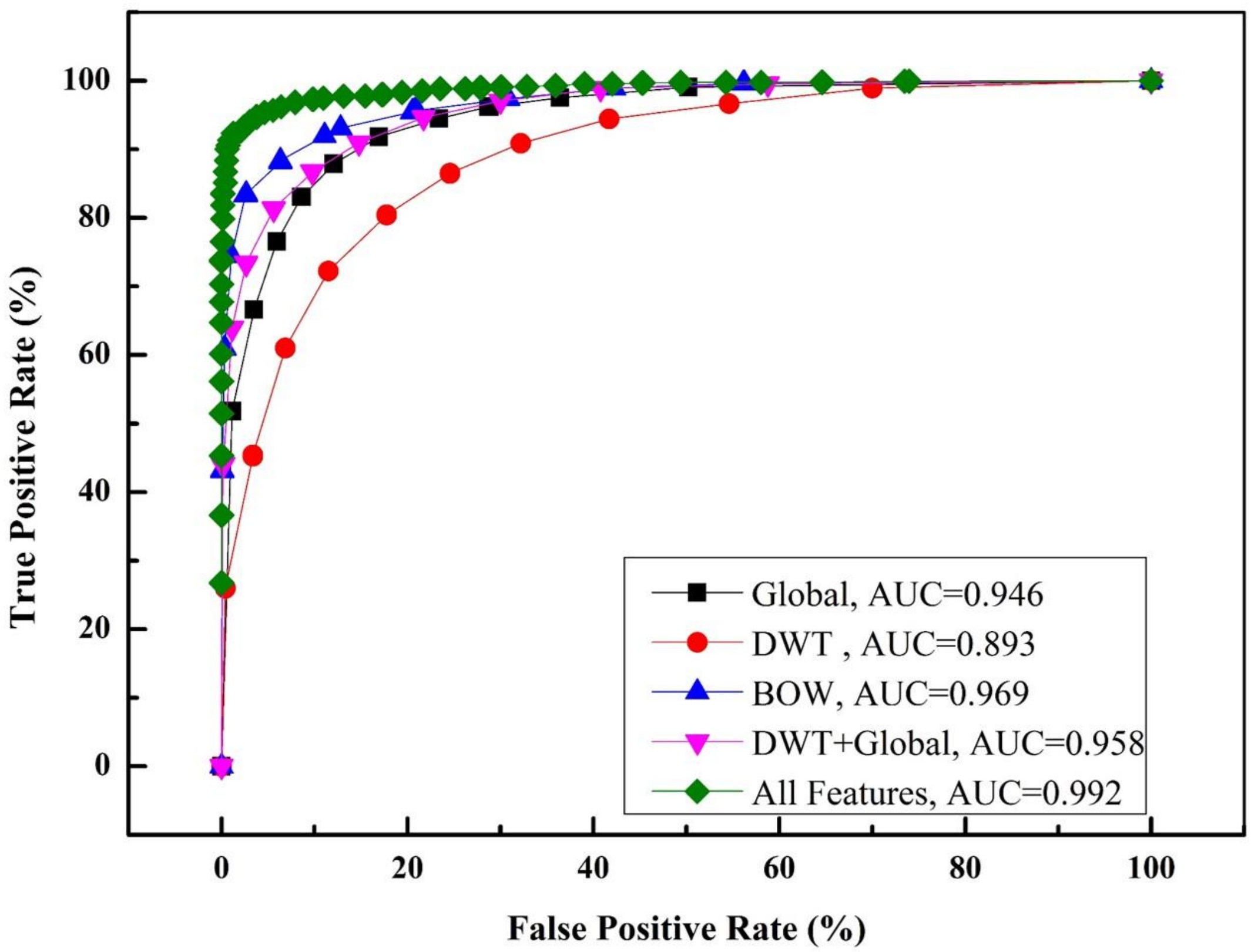

Fig. 11. A comparison of ROC curve of different combination of features.

## 5.7 Comparing with the state-of-the art

In this section, we present some comparison between the best configurations of our method with other existing methods. The best configuration of the method can achieve ROC value of 99.2% using Random Forest (500 trees) and with the local segment size and codebook size are 6 and 250 respectively. We compare our methods with the existing ones from two aspects. One the one hand, we compare our methods with some commercial antivirus software. On the other hand, our method also is compared with some prevalent tools. Overall we collect 1009 malicious documents which include Microsoft Office documents (like RTF, DOC, PPT, DOCX, PPTX, XLS, etc.) and Adobe Reader documents (PDF). The dataset combined malicious documents and benign document, with the number are 512 and 520 respectively. Some of them are the samples used in the APT attacks in recent years, like APT28, APT29 and so on. We extract the hash value of those samples from the analysis report that published by some security vendors, like Fire Eyes. Then we get the related samples on the internet through a hash value. Clean documents are mostly collected from our labs which do not contain any personal privacy information. Finally, we filter out some samples that appear in the training samples through MD5 value of the file. For comparison with the commercial anti-virus software, we selected top 15 leading anti-virus engines. We upload all the sample to VirusTotal to see the results generated by the selected anti-virus engines. We compare the detectors in terms of their true positive rate in the course of our experiments. By definition of our ground truth, most anti-virus software just detects the based on the signature of their database, therefore, there have no false positive about the anti-virus software. Therefore, it cannot be compared in that sense. In fact, in our method, there are 12 clean samples were identified as malicious. Figure. 12 presents the result of $TPR$. Here the most accurate anti-virus engines, Emsisoft and Bitdefender, have the $TPR$ of 0.908 and 0.906 respectively. While our method's obtains the $TPR$ with 96.1% which is better than all the other anti-virus engines.

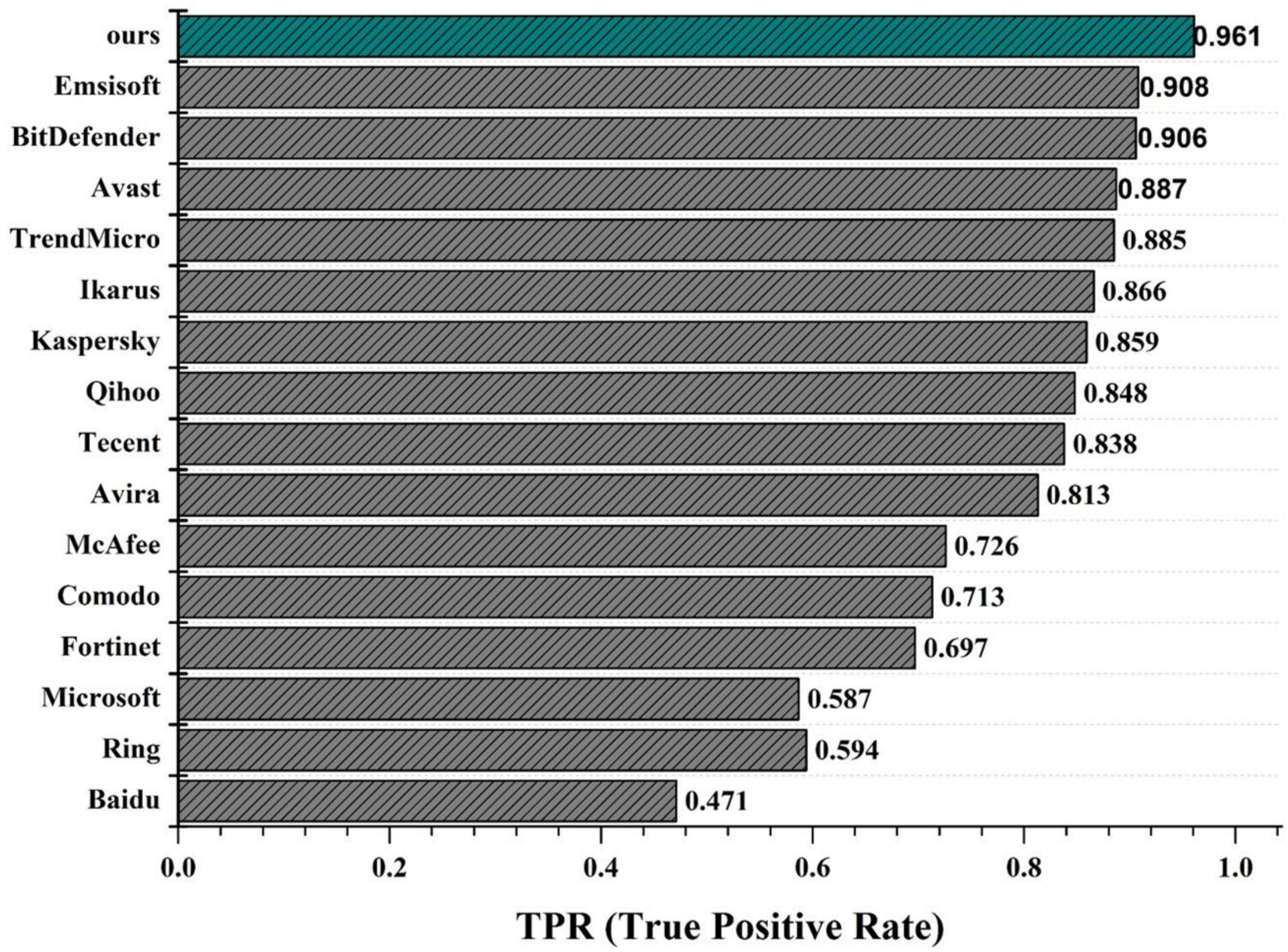

Fig. 12. The TPR of ESRMD compared with top15 leading anti-virus engines.

In addition to the comparison with anti-virus engines, we also compare our method with some prevalent tools which include two methods based on dynamic analysis and static analysis. In this section, we conduct tree groups of comparative experiments to evaluate the efficiency of our models. Table. 4 illustrates the tools we choose to compare with our Framework and the results.

Table. 4. A comparison of ESRMD with other tools

| Features | Approaches | Formats | Time (s) | TPR (%) | PRE (%) | FNR (%) | F1-score (%) |
|---|---|---|---|---|---|---|---|
| Xede [17] | Dynamic | Office & PDF | 152.3 | 93.64 | 95.5 | 4.22 | 94.56 |
| **ESRMD** | **static** | **Office & PDF** | **12.48** | **96.02** | **95.25** | **4.74** | **95.63** |
| PDFrate [13] | static | PDF | 7.95 | 95.94 | 95.34 | 4.05 | 95.63 |
| PDFexaminer [12] | Static & dynamic | PDF | 130.525 | 95.71 | 83.75 | 16.45 | 89.33 |
| **ESRMD** | **static** | **Office & PDF** | **15.96** | **95.45** | **94.02** | **6.25** | **94.72** |
| MalOfficeScanner [11] | static | Office | 2.23 | 83.92 | 97.19 | 3.51 | 90.06 |
| Cryptam [1] | Static & dynamic | Office | 157.02 | 96.67 | 95.14 | 5.92 | 96.04 |
| **ESRMD** | **static** | **Office & PDF** | **10.58** | **98.10** | **96.10** | **5.55** | **97.08** |

In those tools, most of them provide web services for users to upload samples for detection except for MalOfficeScanner. Therefore, we upload the samples to these web sites to get the detection results and scan the samples with the latest version of MalOfficeScanner. For the first group, we compared our method with an excellent tool named Xede. Xede is a comprehensive detection platform. It can detect both code injection and code reuse attacks. It has been demonstrated that it has a good performance by compared with many anti-virus software. We conduct the comparative experiment on samples including Office and PDF formats. In general, ESRMD and Xede perform almost equally on the detection malicious document. But ESRMD has a great advantage in terms of time consumption. It takes an average of 12.48 seconds to detect a malicious document, while Xede takes about 152 seconds. In the second group experiments, we compare ESRMD with some tools which are just used to detect malicious PDF document, including PDFrate and PDFexaminer. PDFRate is a machine learning classifier based on the structural feature of PDF files. PDFexaminer by Malware Tracker is able to scan the PDF for several known exploits, allows the user to explore the structure of the files, as well as examine, decode and dump PDF object content. From the results, we know that ESRMD's performance is slightly inferior to PDFRate, but it performs much better than PDFexaminer. The advantage of ESRMD over PDFRate is that it can detect more types of malicious documents. In the last group, we compared ESRMD with some tools for detecting malicious Office documents, which are MalOfficeSanner and Cryptam. Cryptam can decrypt embedded malware or obfuscated code by using static and dynamic analysis. It supports for combinations of various lengths of XOR encryption, bitwise ROL or ROR shifting, bitwise NOT, and transposition ciphers. OfficeMalScanner is a static analysis tool which is used to scan for malicious trace. The result shows that ESRMD are better than Cryptam in both time consumption and performance. For the comparison with MalOfficeScanner, there is a significant improvement in the performance, and with only a slight deficiency in time. As we can see from the results, it takes longer to detect malicious PDF document. The reason for this is that it takes longer time to preprocess the PDF document. Most objects embedded in PDF document are compressed or encoded. We must decompress or decode those objects before we extract features from the documents. In general, ESRMD has better performance in comparison with other tools. It can detect more formats of malicious documents, spend less time and have higher accuracy.

# 6. Threats to validity

This section describes the threats that can affect the validity of our method, known as: construct, internal and external validity.

## 6.1 Construct validity

A construct validity factor in our research is represented by the restricted samples of our dataset. In order to alleviate this factor we use 3-fold cross-validation, and each experiment was repeated three times using testing data formed by different samples in order to evaluate every sample forming the full dataset.

## 6.2 Internal validity

Threats to internal validity is the approximate truth about inferences regarding cause-effect or causal relationships.

Our result are effected by many parameters, such as machine learning algorithms, size of Local Segment and Codebook, configuration of the selected classifier and the impaction of the features. In order to mitigate those factors, we have conduct different experiments to select the best suitable classifier and the best configuration for the classifier and the parameters for BOW.

## 6.3 External validity

Threats to internal validity is the extent to which the results of a study can be generalized to other situations and to other people.

The code and dataset adopted to run the experiments are available at https://github.com/coolsmurfs/Malicious-document-detectation. A detailed instruction to conduct the experiments has been given. The developed scripts require the Python interpreter.

# 7. Conclusion and future work

In this paper, we mainly investigate the characters of the entropy distribution of the malicious documents. We use different ways to extract the global and structural entropy features that hidden in the entropy distribution of the file. When extracting global features, we extract some statistical information. During the structural entropy features extraction, we use two ways which are DWT and BOW to extract the detail and local features. These features are combined together and then used to train a classifier. We conduct a lot of experiments to evaluate the validity of the model and optimize many parameters. The experiments show that the features extracted from the entropy time series can effectively assist us to detect malicious documents. We also conduct some experiments which compared our method with some leading anti-virus engines and prevalent tools. The result of the experiments show that our approach has a better performance than other tools and anti-virus engines. It can detect more formats of malicious documents, spend less time and have a higher accuracy.

Future work will be head to enhance the robustness and improves the accuracy of the mode. We would like to use Deep learning algorithms to extract features from the time series and train new model to enhance the ability to identify malicious documents.

## Acknowledgements

This work is supported by the National Key Research and Development Plan (No.2017YFB0802900), National Natural Science Foundation of China (No. 61871279), CCF-Venustech HongYan research funding program (No.CCFVenustechRP2017002). The authors also would like to thank the constructive comments from an anonymous reviewer and the associate editor that helped us to improve the manuscript.